\documentclass[12pt]{article}
\def\dif{{\rm d}}
\def\deriv{\@ifnextchar[{\@deriv}{\@deriv[]}}
   \def\@deriv[#1]#2#3{\mathchoice%
{{\dif^{#1}#2\over\dif{#3}^{#1}}}{{\dif^{#1}#2/\dif{#3}^{#1}}}%
+{{\dif^{#1}#2\over\dif{#3}^{#1}}}{{\dif^{#1}#2/\dif{#3}^{#1}}}}

\def\secteqno{\@addtoreset{equation}{section}%
\def\theequation{\thesection.\arabic{equation}}}
\newcounter{subequation}
\def\thesubequation{\alph{subequation}}
\def\sneqnarray{\stepcounter{equation}\let\@currentlabel=\theequation
\setcounter{subequation}{1}
\def\@eqnnum{{\rm (\theequation.\thesubequation)}}
\global\@eqcnt\z@\tabskip\@centering\let\\=\@eqncr\let\@@eqncr=\@@sne
qncr
$$\halign to \displaywidth\bgroup\@eqnsel\hskip\@centering
 $\displaystyle\tabskip\z@{##}$&\global\@eqcnt\@ne
 \hskip 2\arraycolsep \hfil${##}$\hfil
 &\global\@eqcnt\tw@ \hskip 2\arraycolsep $\displaystyle\tabskip\z@{##}$\hfil
  \tabskip\@centering&\llap{##}\tabskip\z@\cr}
\def\endsneqnarray{\@@sneqncr\egroup $$\global\@ignoretrue}
\def\@@sneqncr{\let\@tempa\relax
   \ifcase\@eqcnt \def\@tempa{& & &}\or \def\@tempa{& &}
   \else \def\@tempa{&}\fi
     \@tempa \if@eqnsw\@eqnnum\stepcounter{subequation}\fi
     \global\@eqnswtrue\global\@eqcnt\z@\cr}
\def\nobiblabels{\def\@lbibitem[##1]##2{\@bibitem{##2}}}
\makeatother
\def\ben{\begin{enumerate}}
\def\een{\end{enumerate}}
\def\beq{\begin{equation}}
\def\eeq{\end{equation}}
\def\bea{\begin{eqnarray}}
\def\eea{\end{eqnarray}}
\def\beann{\begin{eqnarray*}}
\def\eeann{\end{eqnarray*}}
\def\beasn{\begin{sneqnarray}}
\def\eeasn{\end{sneqnarray}}

\begin{document}

\begin{titlepage}

\setcounter{page}{0}
\begin{flushright}
Imperial/TP/2-03 /14 \\
\end{flushright}

\vspace{5mm}
\begin{center}
{\Large {\bf Generally covariant theories: the Noether obstruction
for realizing certain space-time diffeomorphisms in phase space}}
\vspace{10mm}

{\large
Josep M. Pons$^{a,b}$ } \\
\vspace{5mm} $^a${\em Departament d'Estructura i Constituents de
la Mat\`eria,
Universitat de Barcelona,\\
Diagonal 647, E-08028 Barcelona, Spain.}\\[.1cm]
$^b${\em Theoretical Physics Group, Blackett Laboratory, Imperial
College London, SW7 2BZ, U.K.} \vspace{5mm}

\vspace{10mm}
\end{center}
\vspace{5mm}

\begin{abstract}

Relying on known results of the Noether theory of symmetries
extended to constrained systems, it is shown that there exists an
obstruction that prevents certain tangent-space diffeomorphisms to
be projectable to phase-space, for generally covariant theories.
This main result throws new light on the old fact that the algebra
of gauge generators in the phase space of General Relativity, or
other generally covariant theories, only closes as a soft algebra
and not a a Lie algebra.

The deep relationship between these two issues is clarified. In
particular, we see that the second one may be understood as a side
effect of the procedure to solve the first. It is explicitly shown
how the adoption of specific metric-dependent diffeomorphisms, as
a way to achieve projectability, causes the algebra of gauge
generators (constraints) in phase space not to be a Lie algebra
---with structure constants--- but a soft algebra
---with structure {\it functions}.

\end{abstract}


\vfill{
 \hrule width 5.cm
\vskip 2.mm {\small \noindent E-mail: pons@ecm.ub.es}}

\end{titlepage}

\newpage




\section{Introduction}

\subsection{Diffeomorphisms in canonical general relativity}

Diffeomorphisms are the gauge symmetries of general relativity
(GR). However, it has been so far impossible to realize the
complete Lie algebra of diffeomorphisms in the canonical formalism
\cite{arnowitt/deser/misner/62} (ADM) of GR. This -in principle-
limitation of the canonical approach raises at least two immediate
and relevant questions: a), one may wonder what the gauge group in
canonical formalism of GR is, given that the diffeomorphism
algebra is not properly realized, and b), one may ask for the
reason that prevents this realization from being obtained.

On the other hand, and related to this fact, it is well known that
the algebra of constraints in the phase space formulation of
general relativity closes as a soft algebra, that is, with
structure functions instead of structure constants, and not as a
Lie algebra as one would have expected. This could also raise a
new question, c), as to whether we realize any group structure at
all for the gauge transformations in phase space.

As we said, these problems have been identified for a long time.
As regards the questions related to the gauge group realization in
phase space and its relationship to the diffeomorphism group
(questions a) and c)), they have been answered some time ago,
\cite{bergmann-komar,Teitelboim:fb,Teitelboim:1972vw}, and can be
given a complete understanding in terms of the concept of a
diffeomorphism-induced gauge group \cite{Pons:1996pr}. What we
think still needs clarification is question b), as to why the
canonical formalism of GR {\sl can not} realize the Lie algebra of
diffeomorphisms. We shall connect this question to that of the
projectability of diffeomorphisms onto phase space.

In the approach taken in \cite{bergmann-komar}, and further
pursued in \cite{Salisbury83,Salisbury83b}, the problem of
realizing diffeomorphisms in phase space was properly addressed in
the following way. The gauge generators in phase space are
constructed with linear combinations of constraints containing
arbitrary functions {\sl and} their first time derivatives. The
Poisson bracket of two of these generators, say $G[\xi_1],
G[\xi_2]$, with $\xi_1$ and $\xi_2$ arbitrary functions of the
space-time variables, should be of the form $G[\xi_3]$ for some
$\xi_3$, but the standard diffeomorphism rule $\xi^\mu_3 =
\xi_2^\nu \xi_{1,\nu}^\mu - \xi_1^\nu \xi_{2,\nu}^\mu $ can not be
implemented by an equal-time commutator, as is the Poisson
bracket, because $\dot\xi_3$, which appears in $G[\xi_3]$, will
depend on the second time derivatives of $\xi_1$ and $\xi_2$. This
dependence can not be generated by the equal-time Poisson Bracket
$\{G[\xi_1], G[\xi_2]\}$.

The main aim of this paper is to give a complementary and more
structural explanation for the same fact from the general
perspective of the Noether theory of symmetries, by pointing out
where the obstruction resides that prevents certain
diffeomorphisms to be projectable to phase space. This result will
emerge as an application of the Noether theory of symmetries
extended to gauge theories\footnote{Here by gauge theories we mean
theories, formulated through a variational principle, containing
symmetries -gauge symmetries- that depend on arbitrary functions.
This includes Yang-Mills theories, string theory, general
relativity etc.}, or more concretely, from the characterisation of
the Noether conserved quantities in phase space.

A second aim of this paper is to show the deep connection between
two issues, a) the one just mentioned concerning the Noether
obstruction to obtain certain diffeomorphisms in phase space, and
b) the well known fact that the algebra of gauge generators in the
phase space of GR only closes as a soft algebra and not a a Lie
algebra. Our analysis will make it clear why this fact is
inevitable and has its roots in the procedure adopted to
circumvent the first problem, which is that of introducing
field-dependent diffeomorphisms in order to achieve
projectability. We may say that it seems very unlikely that there
can be an alternative procedure, within the standard canonical
formalism, to solve the first problem without causing the second
one to appear. Even though we do not claim to solve the problem of
the soft algebra realisation, we think that providing with a
better understanding of its origin may open new ways to solve it
or to prove that such a solution is not possible in the present
framework.

To proceed, we will use the characterization, obtained in
\cite{Pons:1999az}, of Noether symmetries (including both rigid
and gauge symmetries) in the canonical formalism for gauge
theories. So we will first quote the results we need in order to
analyze the case of general relativity or, eventually, other gauge
theories. Although we are interested in gauge field theories, we
will use in this part the language of mechanics, which is
sufficient for our purposes. A quick switch to the field theory
language can be achieved by using DeWitt's \cite{dewitt} condensed
notation\footnote{As we shall see, spatial boundaries will not be
relevant to our discussion, although they are indeed so when one
takes into account that boundary terms can be needed for the
action in order to get a correct formulation of the variational
principle or the conservation of charges.}.

In a series of papers,
\cite{Pons:1996pr,Salisbury:1999rv,Pons:1999ck,Pons:1999xu,Pons:1999xt},
the realization of the diffeomorphism-induced gauge group in phase
space for several generally covariant theories has been studied,
and the projectability issue of diffeomorphisms, which is examined
at the level of the symmetry transformations, is addressed in
detail. In contrast with this previous approach, the Noether
theory for gauge systems will now allow us to directly link the
projectability requirements of diffeomorphisms with specific
properties of their corresponding Noether conserved quantities.
This is one of the novelties of our approach.

It is worth mentioning another approach to addressing the problem
of realizing the diffeomorphism algebra in phase space and
circumventing the difficulties raised above. In the framework
introduced in \cite{Isham:1984sb,Isham:rz}\footnote{See also the
developments in \cite{Halliwell:1990qr} concerning the path
integral formulation within this approach.}, Isham and Kuchar
enlarge the phase space with a set of scalar fields that represent
embedding space-time coordinates that have been promoted to the
status of phase space variables. Remarkably it is then possible to
obtain the diffeomorphism algebra in this augmented phase space.
Let us also mention that in a phase space histories version
\cite{Savvidou:2001dt} of canonical GR, the interpretation of two
``types of time" makes it compatible, by defining a new Poison
bracket in the space of histories, the existence of both the
diffeomorphism algebra structure and the constraints' algebra
structure.

\section{Noether symmetries in gauge theories}

Gauge theories present very specific features concerning the way
the Noether theory of symmetries is implemented. Let us list the
main ones. Consider, as our starting point a time-independent
first-order Lagrangian $L(q,\, \dot q)$ defined in tangent space
$TQ$, that is, the tangent bundle of some configuration manifold
$Q$. The -infinitesimal- Noether symmetries in tangent space we
will consider are of the type $\delta^L q(q,\dot q;t)$. Gauge
theories rely on singular -as opposed to regular- Lagrangians,
that is, Lagrangians whose Hessian matrix with respect to the
velocities ($q$ stands, in a free index notation, for local
coordinates in $Q$ ), \beq W_{ij}\equiv
{\partial^2L\over\partial\dot q^i\partial\dot q^j}, \label{hess}
\eeq is not invertible.

Notice first that these -gauge- theories, having been defined
through singular Lagrangians, allow for the possible existence of
gauge -also called local- symmetries. These symmetries have the
distinctive property of depending on arbitrary functions, and are
deeply connected to certain identities, the Noether identities,
which relate the Euler-Lagrange derivatives of the Lagrangian to
their time -or space-time- derivatives to several orders. Clearly,
these Noether identities are peculiar to singular Lagrangians.

A second feature resides in the special characteristics of the
associated canonical formalism in phase space $T^*Q$. As Dirac
showed \cite{Dirac:pj,dirac} in his pioneering work, dynamics can
be still formulated in phase space but one must take into account
two main novelties, absent in the regular case. First, the
presence of constraints, that is, regions in phase space of low
dimensionality identified as the only places where the equations
of motion may have solutions; consistency of these constraints
with the dynamics is always an important aspect to be analyzed and
solved. And second, the presence of arbitrary functions in the
dynamics as an explicit manifestation of the gauge phenomenon,
which allows for the existence of {\it several} (in fact
infinitely many) dynamical solutions, all starting with the {\it
same} set of initial conditions. All these solutions must be
considered as physically equivalent and are linked by gauge
transformations.

A third feature concerns the very relation between the formalisms
in velocity space and phase space. Since the singularity of the
Hessian matrix makes the usual trade between velocities and
momenta no longer possible, one is led to the issue of the {\it
projectability} -or lack of it- of the structures from velocity
space to phase space. This issue of projectability will be central
in our presentation.

It is clear so far that there are many differences between the
singular case -the one we are interested in- and the regular case
(that is, when the map from velocity space to phase space is
invertible). The extent to which these differences affect the
formulation of Noether symmetries and conserved quantities is
described below.

\vspace{5mm}

Either in the regular or in the singular case, a Noether conserved
quantity $G^L(q, \dot q;t)$ and its associated infinitesimal
transformation $\delta^L q(q, \dot q;t)$ are always linked by the
basic relation
\begin{equation}
[L]_i\delta^L q^i + \frac{d^L}{d \, t}( G^L) = 0 \,,
\label{noet}
\end{equation}
where $[L]_i$ stands for the Euler-Lagrange equations
$$
[L]_i := \alpha_i - W_{is}\ddot q^s\,,
$$
with
$$
\alpha_i :=
    - {\partial^2L\over\partial\dot q^i\partial q^s}\dot q^s
    + {\partial L\over\partial q^i} \,,
$$
and the total time derivative is, in our case, \beq
\frac{d^L}{d\,t} = \frac{\partial}{\partial\,t} + {\dot
q^i}\frac{\partial}{\partial\,q^i} + {\ddot
q^i}\frac{\partial}{\partial\, {\dot q^i}} \, . \label{timeder}
\eeq
\subsection{The regular case}
Now consider a regular theory. In this case, after the trading
between velocities and momenta, we can write the Noether conserved
quantity in phase space, $G^H(q,p;t)$ such that $G^L = {\cal
F}\!L^*(G^H)$ (${\cal
  F}\!L^*$ is the pullback of the Legendre map from $TQ$ to $T^*Q$,
that is, ${\cal F}\!L^*(p) = {\partial L / \partial \dot q}$ ),
and one question may easily come to mind: is there any
characterization of $G^H$ in phase space? the answer is yes: the
canonical conserved quantity satisfies ($H$ is the Hamiltonian)
\beq \frac{\partial \, G^H}{\partial \, t} + \{ G^H,\, H \} = 0,
\label{regular-cond} \eeq if and only if its associated $G^L$
satisfies (\ref{noet}) for some transformations $\delta^L q$. This
transformation $\delta^L q$ can be written as \beq \delta^L q
={\cal F}\!L^* \{ q \, , G^H \} \,, \label{delq-reg}
 \eeq explicitly
showing that $G^H$ is the generator of the canonical Noether
transformation $\delta^H q = \{ q \, , G^H \}$, which is the phase
space version of $\delta^L q = {\cal F}\!L^*(\delta^H q)$. This
classical result is remarkable: a single function, $G^H$, codifies
all the information contained in the Noether symmetry $\delta^H
q$.

Now the obvious question is: how can all these results that hold
in a regular theory be translated, and to what extent, to the
singular case? The precise answer is given in \cite{Pons:1999az},
relying on earlier work done in \cite{Batlle:1985ss,Gracia:af}.

\subsection{The singular -gauge- case}

In the singular case -singular Lagrangians defining gauge
theories-, the conserved quantity $G^L$ in (\ref{noet}) is still a
projectable quantity \cite{kami82}, that is, there exists in phase
space a function $G^H$ such that $G^L = {\cal
F}\!L^*(G^H)$\footnote{An important example of this assertion is
Dirac's canonical Hamiltonian itself, whose
  pullback is the Lagrangian energy, the Noether conserved quantity
  associated with time translations.}.
However, this function $G^H$ is no longer unique because we can
add to it an arbitrary linear combination of primary constraints.
Also, in contrast with the regular case, and despite the existence
of a conserved quantity in phase space, neither $G^H$ nor any of
its equivalent functions whose pull-back to tangent space is
$G^L$, are guaranteed to generate the Noether transformation
$\delta^L q$. By the same token, it is by no means guaranteed that
$\delta^L q$ is a transformation projectable to phase space. These
are the issues we will address next.

Before giving the complete results let us introduce some notation.
The canonical
Hamiltonian will be written $H_c$ and it has the property that its
pullback
to $TQ$ gives the Lagrangian energy, defined as $\dot q \ {\partial L /
\partial  \dot q}  - L$. As we said, this canonical Hamiltonian has
the ambiguity of the possible addition of some functions, the
primary constraints, which we now introduce. The primary
constraints, which will be denoted as $\phi_\mu$ (with the index
$\mu$ running the appropriate values) span a basis for the ideal
of functions in $T^*Q$ whose pullback to $TQ$ under the  Legendre
map ${\cal
  F}\!L$ vanishes, that is, ${\cal F}\!L^*(\phi_\mu)=0$.
Following Dirac, we can split the primary constraints between
those that are first class, $\phi_{\mu_0}$, and the rest, second
class, $\phi_{\mu_1}$, such that,
\begin{equation}
\{ \phi_{\mu_0},\,\phi_{\mu} \} = pc, \quad \det |\{
\phi_{\mu_1},\,\phi_{\nu_1} \}| \not= 0\,, \label{fc-sc}
\end{equation} where $\{-,\,- \}$ is the Poisson Bracket structure
and $pc$ stands for a generic linear combination of the {\sl
primary} constraints. We will assume henceforth that the
determinant in (\ref{fc-sc}) will be different from zero
everywhere in the surface of primary constraints.

Let us also mention that the vector field generating the dynamics
(time evolution) in the canonical formalism for gauge theories
(also called {\sl constrained systems} in the Dirac's sense
\cite{Dirac:pj,dirac}) is:
\begin{equation} {\bf
X_H}:={\partial \over \partial t} + \{-,\,H_c \} + \lambda^\mu
\{-,\, \phi_\mu \}\,, \label{xh}
\end{equation}
where $\lambda^{\mu}$ are arbitrary functions of time. Consistency
of the dynamics (\ref{xh}) with the primary constraints leads
generally to new constraints and to new refinements of the
dynamics (see \cite{Batlle:1985ss}).

It is very convenient at this point to realize that the functions
$\lambda^{\mu}$, despite being, at this moment, arbitrary
functions in phase space, admit an unambiguous determination in
velocity space, namely, the functions $v^\mu(q, \dot q)$ that
satisfy the identity
\begin{equation} \dot
q \equiv {\cal F}\!L^* \{q, \, H_c  \} + v^\mu(q, \dot q) {\cal
F}\!L^* \{q, \, \phi_\mu\} \,. \label{id1}
\end{equation}
These functions $v^\mu$ are strictly non projectable to phase
space. They form in fact \cite{Garcia:2000yt} a basis for the
functions that are not projectable to phase space. They will play
an important role shortly.

It is proved in \cite{Pons:1999az} that the phase space
characterization of $G^H$ for gauge theories is as follows:

\vspace{4mm}

\newtheorem{guess}{Theorem}
\begin{guess}[Garc\'{i}a-Pons, 2000]

The pullback $G^L$ --in $TQ$-- of a function $G^H$ in
$T^*Q$ satisfies (\ref{noet}) --for some $\delta q^L$-- if and
only if $G^H$ satisfies
\begin{equation}
  {\partial G^H\over\partial t} + \{G^H, \,
H_c\}^* =  sc + pc \,, \qquad \{G^H,\, \phi_{\mu_0} \}^* = sc + pc
\,,
 \label{cond}
\end{equation}  where sc (pc) represents a generic combination
of secondary (primary) constraints. \end{guess}

\vspace{5mm}

These secondary constraints are obtained by requiring the
consistency of the dynamics with the primary constraints
(essentially, by requiring tangency conditions of the evolutionary
vector field with respect to the surface of primary constraints;
this requirement may also determine some of the arbitrary
functions $\lambda^\mu$ in (\ref{xh}). Search for constraints may
continue with tertiary constraints, etc.). A basis --perhaps
redundant, perhaps even void-- of secondary constraints can be
written as \beq \phi^1_{\mu_0} := \{ \phi_{\mu_0}, \, H_c\} .
\label{pbconstr} \eeq

The Dirac bracket in (\ref{cond}) is defined, at this level of the
surface of primary constraints, by
$$
\{A,\,B\}^* := \{A,\,B\} - \{A,\,\phi_{\mu_1}\}M^{{\mu_1}{\nu_1}}
\{\phi_{\nu_1},\,B\} \,,
$$
where $M^{{\nu}_1{\mu}_1}$ is the matrix inverse of the Poisson
bracket matrix of the primary second-class constraints (see
(\ref{pbconstr})).

One can notice that (\ref{cond}) is insensitive -as it must be- to
the ambiguity inherent in the definition of $G^H$, that is, under
the addition to a given $G^H$ of linear combinations of primary
constraints. We can rewrite (\ref{cond}) using a specific notation
for the coefficients in the secondary constraints: \bea {\partial
G^H\over\partial t} + \{G^H, \, H_c\}^* = A^{\mu_0} \phi^1_{\mu_0}
+ pc , \nonumber \\
\{G^H, \, \phi_{\mu_0} \}^* = B^{\nu_0}_{\mu_0}\phi^1_{\nu_0} + pc
\,. \label{cond-coef} \eea

The coefficients $ A^{\mu_0}$ in (\ref{cond-coef}) can be absorbed
through a suitable choice of $G^H$ among all the functions whose
pullback is $G^L$; specifically, under the change $G^H \rightarrow
G^H - A^{\mu_0} \phi_{\mu_0}$ we make the new $ A^{\mu_0}$ to
vanish. A new change, suggested in \cite{dirac} as the definition
of a ``starred'' function,
$$
G^H \rightarrow G^H  -
\{G^H,\,\phi_{\mu_1}\}M^{{\mu_1}{\nu_1}}\phi_{\nu_1} \,,
$$
makes irrelevant the use of the Dirac bracket in favor of the
usual Poisson bracket. Notice though that under these changes we
have eliminated, partially at least, the ambiguity inherent in the
definition of $G^H$, so the mathematical characterization for {\sl
the most general} $G^H$ whose pullback is a Noether conserved
quantity, $G^L$, in velocity space, is still (\ref{cond}) (or
(\ref{cond-coef})). Notice finally that the coefficients
$B^{\nu_0}_{\mu_0}$ remain invariant under the changes of $G^H$
allowed by the addition of arbitrary linear combinations of the
primary constraints. These quantities $B^{\nu_0}_{\mu_0}$ play a
fundamental role in what follows.

\subsection{Reconstructing the Noether transformation and finding
the Noether obstruction in phase space}

As we shall now see, these coefficients $B^{\nu_0}_{\mu_0}$ bear
the full responsibility -if they do not vanish- for some Noether
transformations not being projectable to phase space. Indeed one
can prove \cite{Pons:1999az} that

\vspace{4mm}\begin{guess}[Garc\'{i}a-Pons, 2000]
 The reconstruction of the Noether
transformation\footnote{There is a small ambiguity in $\delta^L
q$, as can be seen in (\ref{noet}), because we can add to
$\delta^L q$ an arbitrary antisymmetric linear combination of the
Euler-Lagrange derivatives, whereas (\ref{noet}) keeps unchanged.
Usually these additions introduce accelerations in the new
$\delta^L q$, not allowed in our framework. Let us also point out
that in some cases there can even be identically vanishing Noether
conserved quantities associated with non-trivial gauge Noether
transformations \cite{Gracia:1992pe}. Examples are the
relativistic free particle --without auxiliary variable-- and the
Nambu-Goto action for the string.} $\delta^L q$, out of a function
$G^H$ satis\-fying (\ref{cond-coef}) goes as follows:
\begin{equation}
 \delta^L q = {\cal F}\!L^*(\{q,\,G^H -
A^{\mu_0} \phi_{\mu_0} \}^*) - v^{\mu_0} {\cal F}\!L^*
(B^{\nu_0}_{\mu_0}\{q,\, \phi_{\nu_0} \}^* )\,. \label{delq}
\end{equation}
\end{guess}  This result is the extension to gauge theories of
the simpler relation (\ref{delq-reg}), derived for the regular
case.

\vspace{5mm}

It is clear that in (\ref{delq}) we completely identify the
obstruction that may prevent the projectability of $\delta^L q$ to
phase space, namely, the existence of the second term on the left
hand side. Indeed, when any of the coefficients
$B^{\nu_0}_{\mu_0}$ are different from zero, the functions $
v^{\mu_0}$, which are intrinsically non-projectable, prevent
$\delta^L q$ from being projectable.

Summarizing, given a Noether conserved quantity in tangent space,
$G^L$, we have a direct procedure to construct its associated
Noether transformation and to check whether it is projectable to
phase space: first write any function $G^H$ in $T^*Q$ such that
its pullback is $G^L$, $G^L = {\cal F}\!L^*(G^H)$ (the existence
of such a $G^H$ is guaranteed, and in practice it is not difficult
to find one). Next use the relations (\ref{cond}), compulsory for
our $G^H$, to obtain the coefficients $A^{\mu_0}$ and
$B^{\nu_0}_{\mu_0}$ that allow for the reconstruction of $\delta^L
q$. When the $B$ coefficients all vanish the transformation
$\delta^L q$ will be projectable. Otherwise it will not be.

\section{Application to Noether gauge symmetries: the Maxwell theory}

The most important examples of gauge systems in theoretical
physics are Yang-Mills theory, general relativity (GR) and string
theory. These cases share some features that allow to treat them
together with regard to the construction of their respective
Noether gauge symmetries. In fact, for our purposes, the essence
of Yang-Mills theory is already present in the Maxwell theory,
which is the case we will consider henceforth. We will limit
ourselves to comparisons between Maxwell theory and
GR\footnote{String theory in the
Brink-DiVecchia-Howe-Deser-Zumino-Polyakov formalism can be
essentially treated along the same lines.}.

Gauge transformations in Maxwell theory -$U(1)$ symmetry- and in
general relativity -active space-time diffeomorphisms- share the
fact that the infinitesimal gauge transformation depends on
arbitrary functions (a scalar function $\Lambda$ in
electromagnetism, the components $\alpha^{\mu}$ of a vector field
$\alpha^{\mu} \partial_{\mu}$ in general relativity) of space-time
{\it and} their first space-time derivatives. The particular fact
that the first time derivative of the arbitrary functions is
necessarily present in both cases has direct implications for the
structure of constraints of these theories: there must be
secondary constraints, in addition to the primary ones. This point
will be clarified below. The similarities do not stop there: both
theories exhibit only primary and secondary constraints, and all
of them are first class.

Since all constraints are first class, the Dirac bracket coincides
with the Poisson bracket, so in the cases we are considering,
equations (\ref{cond-coef}) will be read with the Poisson bracket.
It is easy to show that {\it every} secondary constraint (remember
that all constraints are first class) will provide us with a
Noether gauge transformation. Indeed, in order to generate a
transformation depending on an arbitrary function, say
$\epsilon^{\mu_0}$, let us attach it to the secondary constraint:
$\epsilon^{\mu_0}\phi^1_{\mu_0}$. We can just sum over ${\mu_0}$
and thus describe the whole set of gauge transformations. This
object, $G^H:= \epsilon^{\mu_0}\phi^1_{\mu_0}$, satisfies
(\ref{cond}) trivially. The first condition is satisfied because
we know that there are no tertiary constraints in this case and
the second one because all constraints are first class. So we can
plug our $G^H$ into (\ref{cond-coef}) to compute the coefficients
$A^{\mu_0}$ and $B^{\nu_0}_{\mu_0}$ in order to build the gauge
transformations (\ref{delq}).

\subsection{Maxwell theory}

In the case of pure ${\cal E}{\cal M}$, from the Lagrangian
$${\cal L}_M = -\frac{1}{4} F_{\mu \nu}F^{\mu \nu} \,,
$$
we get the canonical Hamiltonian
$$H_c = \int d{\bf x} \left[\frac{1}{2}({\vec \pi}^2
+ {\vec B}^2) + {\vec \pi} \cdot \nabla A_0\right] \,,
$$
and a primary constraint $\pi^0 \simeq 0$. Stability of this
constraint under the Hamiltonian dynamics leads to the secondary
constraint $\dot\pi^0 = \{\pi^0, H_c\} = \nabla \cdot {\vec \pi}
\simeq 0$. Both constraints are first-class and no more
constraints arise. According to the considerations above, it is
immediate to write down the quantity $G^H$ that satisfies
(\ref{cond}),
$$
G^H[\Lambda ; t] = \int d^3\!x \, \Lambda({\bf x},t) \,\,\nabla
\cdot {\vec \pi}({\bf x}, t) \,,
$$
$ \Lambda$ being the arbitrary gauge function. One readily
determines the quantities $A$ in (\ref{cond-coef}) and realizes
that the quantities $B$ vanish. Therefore the gauge transformation
is projectable to phase space and canonically generated -through
the Poisson bracket- by
$$
G[\Lambda ; t] = \int d^3\!x \, \left[-\dot{\Lambda}({\bf x}, t)
\pi^0({\bf x}, t) + \Lambda({\bf x},t) \nabla \cdot {\vec
\pi}({\bf x}, t)\right] \,.
$$

The gauge transformation of the gauge field is then
$$\delta A_{\mu} =
\{A_{\mu}, \, G\} = -\partial_\mu{\Lambda} \,,
$$
which is the usual Noether $U(1)$ symmetry for the Lagrangian
${\cal L}_M$. Let us observe that a primary and a secondary
constraint are necessary to ensure that the gauge field $A_{\mu}$
transforms covariantly.

\section{General Relativity}

One could obviously write four gauge transformations, one for each
secondary constraint, following the procedure used for ${\cal
  E}{\cal M}$.
But that will not directly answer our question as to whether
diffeomorphisms in tangent space are projectable to phase space.
To this end we should rather start by constructing the conserved
quantity $G^L$ associated with diffeomorphisms in tangent space,
obtaining its phase space version $G^H$ and then checking whether
the conditions of projectability --the vanishing of the
coefficients $B^{\nu_0}_{\mu_0}$ in (\ref{cond-coef})-- are met.
The most efficient way to get $G^L$ in general relativity for the
pure gravity case is by making use of the doubly contracted
Bianchi identities, which are the geometric version of the Noether
identities. Let us see how it works.

Now we will use the language of field theory. Consider a
Lagrangian density ${\cal L}$, with some fields\footnote{Unless
stated otherwise, internal or space-time indices for each field
will not be displayed.} $\psi_A$ and having the Noether gauge
symmetry \beq \delta \psi^A = \epsilon f^A +  (\partial_\mu
\epsilon) f^{A \mu} \,, \label{gaugesymm} \eeq for an arbitrary
function $\epsilon$ of space-time and for given functions $f_A, \,
f_A^\mu$, of the fields and their first space-time derivatives.
Extension to higher space-time derivatives of the arbitrary
function is straightforward but it is not needed here.

Use of the Noether condition (\ref{noet}), and the fact that
 the infinitesimal function $\epsilon$ is arbitrary, produces
the Noether identity ($[{\cal L}]_A$ stands for the Euler-Lagrange
derivatives of ${\cal L}$), \beq [{\cal L}]_A f^A = \partial_\mu
([{\cal L}]_A f^{A\mu}) \,, \label{noet-id} \eeq from which we can
obtain \beq [{\cal L}]_A \delta \psi^A = \partial_\mu (\epsilon
[{\cal L}]_A f^{A\mu}) \,, \label{curr} \eeq which identifies the
conserved current as an object vanishing on shell, $J^\mu :=
\epsilon [{\cal L}]_A f^{A\mu}$. Notice that (\ref{gaugesymm}) and
(\ref{noet-id}) are connected in both ways: one can either derive
the Noether identity (\ref{noet-id}) from the gauge transformation
(\ref{gaugesymm}) or vice-versa, construct the gauge
transformation out of the Noether identity.

Let us apply these ideas to general relativity. The Einstein
tensor density,
$$
{\cal G}_{\mu\nu} :=\sqrt{|g|}(R_{\mu\nu} - {1 \over 2} R
g_{\mu\nu}) \,,
$$
satisfies the Bianchi identities \beq {\cal G}^{\mu\nu}_{;\nu} =0
\,. \label{bianchi} \eeq

Now, in the case of pure gravity the Euler-Lagrange derivatives of
the Einstein-Hilbert Lagrangian are just the components of the
Einstein tensor density,
$$[{\cal L}_{EH}]^{\mu\nu} = {\cal
G}^{\mu\nu}\,,$$ and the content of (\ref{bianchi}) becomes that
of a Noether identity. Indeed it can be equivalently expressed in
the form of (\ref{curr}), \beq {\cal G}^{\mu\nu} \delta
g_{\mu\nu}=
\partial_\rho (2 \epsilon^\lambda {\cal G}_\lambda^{\ \rho})\,,
\label{bianchi-noether}
\eeq
with $\delta g_{\mu\nu} =
\epsilon^\rho \partial_\rho g_{\mu\nu} + g_{\mu\rho} \partial_\nu
\epsilon^\rho + g_{\rho\nu} \partial_\mu \epsilon^\rho$, that is,
the infinitesimal diffeomorphism -the Lie derivative- generated by
the vector field $\epsilon^\rho \partial_\rho $ . We recognize in
(\ref{bianchi-noether}) the Noether conserved current under the
gauge symmetry of diffeomorphisms
$$
J^\rho := 2 \epsilon^\lambda {\cal G}_\lambda^{\ \rho} \,.
$$

It is well known that there is an intrinsic ambiguity in the
definition of the current $J^\rho$: a change of the type \beq
J^\rho \rightarrow J^\rho +
\partial_\mu A^{\mu\rho} \label{bterms} \eeq with any
antisymmetric $A^{\mu\rho}$ leaves (\ref{bianchi-noether})
invariant. The space integration of the time component $J^0$ is
the putative Noether conserved charge $G^L = \int d^3\!x J^0$.
Such conservation relies on the vanishing of the flux of the
3-vector $J^{i}$ through the spatial boundary. Since $J^0$
vanishes on shell, it is clear that if there are to be any
non-trivial conserved quantities at all, they must come about by
way of the boundary terms afforded by (\ref{bterms}). Here we see
the relevance of the ambiguity in $J^\rho$, for it could be
possible in some cases to adjust an $A^{\mu\rho}$ piece in such a
way that the new $G^L$ -which is the old one plus boundary terms-
becomes a truly --and non-vanishing-- conserved
charge\footnote{The paramount example of this observation is the
ADM energy.}. All the same these considerations are not important
for our present purposes, because $J^0$ is already a constraint
(in the Dirac sense) and hence $G^L$ is trivially conserved -it
vanishes on shell.

The components appearing in  $J^0$ are the well known secondary
constraints of pure GR. Using the Lapse and Shift functions, $N
=:N^0$ and $N^i$ respectively, and following the conventions of
\cite{Wald:rg}, they take the form
\begin{equation}
2 {\cal G}_0^{\ 0} = N^\mu {\cal H}_\mu , \qquad 2 {\cal G}_i^{\
0} = {\cal H}_i \, ,
\end{equation} (here we have used a
``covariant-like'' notation, with indices $\mu = (0, i)$, just to
express summation on the repeated indices, not to imply covariant
behavior) where ${\cal H}_0$ is the Hamiltonian constraint and
${\cal H}_i$ are the momentum constraints. Notice that to isolate
${\cal H}_0$ one essentially uses the unitary vector $n^\mu$
perpendicular to the equal-time foliation,
$$ n^\mu = ({1 \over N},\,  - {N^i \over N}), \quad \ n_\mu =
g_{\mu\nu} n^\nu = ( -N,\, \vec{0}\ )\,,
$$
in the following way
\begin{eqnarray}
-N {\cal H}_0 &=& -2 {\cal G}_0^{\ 0} + 2 N^i{\cal G}_i^{\ 0} = 2
{\cal G}_0^{\ 0}n^0 n_0 + 2 {\cal G}_i^{\ 0}n^i n_0 = 2 {\cal
G}_{\mu}^{\nu}n^\mu n_\nu \nonumber \\ &=& 2 {\cal
G}_{\mu\nu}n^\mu n^\nu = 2 N \sqrt{\det{g_{ij}}} \ ({}^{(3)}\!R +
(K^i_{\ i})^2 - K_{ij} K^{ij}) \nonumber \,,
\end{eqnarray}
thus giving the standard definition \cite{Wald:rg} for the
Hamiltonian constraint ${\cal H}_0$, with $K_{ij}$ being the
extrinsic curvature of the equal-time surfaces.

 Let us observe that \beq H_c := \int d^3\!x \
2{\cal G}_0^{\ 0} = \int d^3\!x \ N^\mu {\cal H}_\mu \,,
\label{canham} \eeq
 is the canonical
Hamiltonian (again, up to boundary terms that do not affect our
discussion), that is, the Noether conserved charge when
$\epsilon^\rho\partial_\rho = \partial_0$. Since the primary
constraints of GR are just the momenta $P_\mu$ conjugate to the
Lapse and Shift, note that the secondary constraints are
$$
\{P_\mu, \, H_c \} = - {\cal H}_\mu \,.
$$

\vspace{4mm}

\underline{\bf Remark.} \ In order to avoid any confusion with the
standard literature, let us observe that, usually, many
presentations of the canonical formalism for GR ignore the primary
constraints $P_\mu$. In such a case, the Lapse and Shift variables
take on the role of Lagrange multipliers for a ``Dirac"
Hamiltonian (\ref{canham}), where the constraints ${\cal H}_\mu$
are taken as ``primary" rather than ``secondary". Although this
procedure is not incorrect, it fails to provide one with the full
variables in phase space -in our case, {\it all} the components of
the metric tensor and {\it all} their canonically conjugate
momenta. In our procedure, instead, we consider that the true
Dirac Hamiltonian is obtained by the addition to (\ref{canham}) of
a combination of the primary constraints with arbitrary Lagrange
multipliers, that is,
\beq H_D := H_c +  \lambda^\mu P_\mu \,.
\label{dirham} \eeq

Obviously the dynamics will impose $\dot N^\mu = \{N^\mu, H_D
\}=\lambda^\mu$ and so we eventually recover, as we should, the
arbitrary character of the Lapse and Shift. Notice, though, the
advantage that we keep the phase space treatment complete,
including Lapse and Shift and their canonical conjugates as
ordinary variables, at every stage. Keeping the phase space
treatment complete means that we do not lose the rules of
transformation for the Lapse and Shift variables (or,
equivalently, for the components $g^{\mu 0}$ of the metric). In
this way, the connection between the gauge symmetries defined in
phase space and the diffeomporphisms defined in tangent space can
be made in very simple terms. Indeed the developments we undertake
below are possible because we keep in the canonical formulation
the entire configuration space of the Lagrangian formalism.

{\bf ---------------} \vspace{6mm}

Now, the conserved charge associated with diffeomorphism
invariance, genera\-ted by the infinitesimal vector field
$\epsilon^\mu
\partial_\mu$, is
$$
G^L = \int d^3\!x J^0 =\int d^3\!x \ (\epsilon^0 N^\mu {\cal
H}_\mu + \epsilon^i {\cal H}_i) \, .
$$

Taking into account the specific form of the Hamiltonian and
momentum constraints, it is immediate to realize that they are
trivially projectable to phase space. Therefore, with the
understanding that we are now expressing our ${\cal H}_\mu$ in
terms of the canonical variables, we can directly write down the
canonical quantity (up to the addition of primary Hamiltonian
constraints, whose pullback to tangent space identically vanishes)
that satisfies (\ref{cond}): \beq G^H = \int d^3\!x J^0 =\int
d^3\!x \ (\epsilon^0 N^\mu {\cal H}_\mu + \epsilon^i {\cal H}_i)
\, . \label{ham-cons-quan} \eeq

Now it is immediate to verify that the first piece in
(\ref{ham-cons-quan}), that is, the conserved quantity $\int
d^3\!x (\epsilon^0 N^\mu {\cal H}_\mu)$ associated with time
diffeomorphisms, produces $B$ pieces in the right side of
(\ref{cond-coef}), thus implying that the time diffeomorphisms are
not projectable onto phase space. The obvious reason is the
presence of the Lapse and Shift functions in this conserved
quantity, which hit in the Poisson bracket with their conjugate
variables, the primary constraints,
$$
\{ \int d^3\!x \ \epsilon^0 N^\mu {\cal H}_\mu , \, P_\nu \} =
\epsilon^0 {\cal H}_\nu \, ,
$$
thus identifying $B^\mu_\nu = -\delta^\mu_\nu \epsilon^0$ .

Once the problem has been identified, we will focus on three
issues. First, we will explain why this problem is unavoidable for
generally covariant theories containing more than scalar fields;
second, we will show the way out, which will of course make
contact with standard formulations; and third, we will observe
that this way out comes at a price: that the algebra of
diffeomorphisms will not be realized as a Lie algebra, but as a
``soft" algebra.
\subsection{The reason for the non-projectability}

We have seen that this non-projectability just happens. Now we
will give a general argument explaining why this must be so.

The crucial observation is that the current in (\ref{curr})
depends only on the arbitrary function $\epsilon$, and so does its
time component $J^0$ (should the current in (\ref{curr}) depend on
higher derivatives of $\epsilon$, then $\delta \psi_A$ would not
be a diffeomorphism transformation of a tensor field). Since its
associated conserved quantity is directly projectable to phase
space, we can write $G^H = \int d^3\!x J^0$ as the quantity that
fulfills (\ref{cond}). Then, use of (\ref{cond-coef}) and
(\ref{delq}) dictates the reconstruction of the diffeomorphism
transformations out of $G^H$.

Now let us concentrate on time diffeomorphisms, and use the
notation $G^H_0$ for its associated conserved quantity. Observe
that $G^H_0$ only contains the arbitrary function $\epsilon^0
:=\epsilon$, and {\sl not} its time derivative $\dot \epsilon$
(space derivatives may be hidden behind spatial integration by
parts). But to reconstruct a time diffeomorphism for the metric
components, we need this time derivative (the only case when it is
not needed is for the transformation of scalars). Where does this
time derivative appear in (\ref{delq})? It can only appear through
the $A$ terms. Thus we conclude that some $A$ terms must be
different from zero, which means that the theory we are
considering, GR, must have secondary constraints.

This is the first step: we have deduced the existence of secondary
constraints. Nothing new, of course, but the point is that this
result has now been obtained from symmetry considerations alone.

Now comes the second step, which concerns the deep relationship
between the conserved quantity $G^H$ for time diffeomorphisms and
Dirac's canonical Hamiltonian $H_c = \int d^3\!x \ {\cal H}_c$. We
do not need to specify ${\cal H}_c$, but simply to realize that
$J^0 = \epsilon {\cal H}_c$. The reason is that when $\epsilon$ is
chosen to be an infinitesimal constant $\delta t$, then time
diffeomorphisms become rigid time translations, and the conserved
quantity associated with time translations is the canonical
Hamiltonian \footnote{Let us stress, however, that despite the
mathematical identification of time evolution with the symmetry of
rigid time translations, both operations substantially differ
\cite{dompep} in their physical meaning: in the active view of
diffeomorphism invariance --which is the one implicitly adopted
thoroughout this paper--, time translations move a trajectory into
another while {\it preserving the value of their time coordinate}
(it is an equal-time operation) --or their space-time coordinates
for field configurations, whereas the infinitesimal time evolution
builds a single trajectory from time $t$ to $t + \delta t$.}.

Now we are ready to get to the final point. Since there are secondary
constraints in the theory, they must be
produced through the Poisson bracket of the canonical Hamiltonian with
the primary constraints.  Schematically,
$$
sc = \{ pc,\, H_C \} \,,
$$
but then it is inescapable that the bracket $\{ pc,\, G^H_0 \}$, with
$G^H_0 = \int d^3\!x J^0 = \int d^3\!x \ \epsilon {\cal H}_c$,
will develop also pieces with secondary constraints, thus producing
$B$ terms in (\ref{cond-coef}) and hence leading to nonprojectability.

This proof of the nonprojectability of time diffeomorphisms to
phase space applies to any generally covariant theory containing
other than scalar fields.
\subsection{The way out}
The nonprojectability of time diffeomorphisms introduces a
potentially damaging problem for the formulation of GR in phase
space, for it could imply that the full contents of diffeomorphism
invariance can not be captured by the canonical formalism. That
this is not true was shown in \cite{Pons:1996pr}, where a
discussion on the gauge group can be found. Here we will show a
direct route to the same result.

Let us start by making a second look at our Noether conserved
quantity for space-time diffeomorphisms, (\ref{ham-cons-quan})
\beq G^H = \int d^3\!x \ (\epsilon^0 N^\mu {\cal H}_\mu +
\epsilon^i {\cal H}_i) =  \int d^3\!x \ \left( \epsilon^0 N^0
{\cal H}_0 + (\epsilon^0 N^i + \epsilon^i) {\cal H}_i \right)\, .
\label{explicitgh} \eeq

Up to now, we have considered $\epsilon^\mu$ as arbitrary
functions of space-time. This means that we have not fully
exploited the arbitrariness of these functions in order to build
actions of the gauge group of GR. Using an arbitrary
$\epsilon^\mu(x)$ (here $x$ represents the space-time coordinates,
$x=(t,{\bf x})$), we produce an infinitesimal action of the gauge
group such that {\sl all} field configurations (the space of field
configurations is the natural arena for the action of the gauge
group) undergo the {\sl same} diffeomorphism. It is obvious that
one can also consider the case where {\sl different} field
configurations can undergo {\sl different} diffeomorphisms under
the action of a single element of the gauge group\footnote{We
refer to \cite{Pons:1996pr} for further considerations on the
gauge group.}. This can be achieved by allowing $\epsilon^\mu$ to
have an arbitrary dependence not only on the space-time
coordinates, but on the field configurations as well. With a
specific selection of these dependences, which amounts to a change
of basis for the infinitesimal generators of the gauge group, it
turns out that we can achieve projectability.

In fact, let us make the functions $\epsilon^\mu$ depend on the
Lapse and Shift in such a way that any dependence on these
variables disappears from (\ref{explicitgh}). That is, we require
that
$$
\epsilon^0(x,N)  N^0 = \xi^0(x), \quad \quad \epsilon^0(x,N) N^i +
\epsilon^i(x,N) = \xi^i(x) \,,
$$
for some functions $\xi^\mu(x)$ that only depend on the space-time
coordinates. Inversion of these relations gives
\beq \epsilon^\mu
= n^\mu \xi^0 + \delta^\mu_i \xi^i \,,
\label{epsilon-xi} \eeq
where $ n^\mu $ is the unitary vector orthogonal to the equal-time
surfaces, introduced before.

Now $G^H$ simplifies to
\beq G^H = \int d^3\!x \ ( \xi^0  {\cal
H}_0 +
 \xi^i {\cal H}_i )\, ,
\label{explicitgh-2} \eeq and it produces vanishing $B$ functions
in (\ref{cond-coef}). We have achieved projectability, but it
comes at a price.

\subsection{The price}

The standard diffeomorphism algebra, for vectors
$ \vec{\epsilon_1} =  \epsilon^\mu_1 (x)\partial_\mu \, , \
\vec{\epsilon_2} = \epsilon^\mu_2 (x)\partial_\mu $  , is
that of Lie derivatives:
$\epsilon^\mu_3 = \epsilon^\nu_2 \epsilon^\mu_{1,\nu}
    - \epsilon^\nu_1 \epsilon^\mu_{2,\nu}
    = [ \vec{\epsilon_2}, \, \vec{\epsilon_1}]^{\mu} $. But this
    is only valid when the functions $\epsilon^\mu_1,
    \epsilon^\mu_2$ depend exclusively on the space-time
    coordinates. If $\epsilon^\mu_1, \epsilon^\mu_2$ are of the
    form (\ref{epsilon-xi}), then $\epsilon^\mu_3$ becomes
$$
\epsilon^\mu_3 = \epsilon^\nu_2 \epsilon^\mu_{1,\nu}
    - \epsilon^\nu_1 \epsilon^\mu_{2,\nu} + \xi_2^0{\cal L}_{\vec{\epsilon_1}} (n^\mu)
- \xi_1^0{\cal L}_{\vec{\epsilon_2}} (n^\mu) \,,
$$
where ${\cal L}_{\vec{\epsilon}} (n^\mu)$ is the Lie derivative of
the ``vector" $n^\mu$. But $n^\mu$ is not a true vector, for it is
constructed algebraically out of the $g^{0\mu}$ components of the
metric tensor. Indeed its transformation rules under the Lie
derivative are \cite{Pons:2001vx}
\beq {\cal L}_{\vec{\epsilon}}
(n^\mu) = {\cal L}_{\vec{\epsilon}}^{\rm (naive)} (n^\mu) + N
h^{\mu\nu}
\partial_\nu\epsilon^0  \, ,
\label{deviation} \eeq with $h^{\mu\nu } := g^{\mu\nu } + n^\mu
n^\nu$ (we take the signature of $g$ ``mostly plus''), and
$${\cal L}_{\vec{\epsilon}}^{\rm
(naive)} (n^\mu) = [\vec{\epsilon}, \, \vec{n}]^\mu \,, $$  is the
``naively" expected vector behavior for $n^\mu$.

\vspace{4mm}

Now, expressing $\epsilon^\mu_3$ as in (\ref{epsilon-xi}), we
obtain
\begin{equation}
\xi_3^\mu = \xi^i_2\xi^\mu_{1,i} - \xi^i_1\xi^\mu_{2,i}
    + h^{\mu\nu }(\xi^0_2 \xi^0_{1,\nu} - \xi^0_1 \xi^0_{2,\nu}) \,,
        \label{comrel}
\end{equation}

These equations already contain the Poisson bracket algebra of
(\ref{explicitgh-2}), that is, \beq \{ G^H[\xi_1], \, G^H[\xi_2]
\} = G^H[\xi_3] \,, \label{thepb} \eeq out of which we can readily
obtain the algebra of the first-class constraints ${\cal H}_\mu$,
$$
  \{{\cal H}_\mu, {\cal H}_\nu \}
  = C^\sigma_{\mu\nu}{\cal H}_\sigma \,,
$$
with the structure functions $C^\sigma_{\mu\nu}$ being determined
by (\ref{comrel}) and (\ref{thepb}).

The presence of the pieces $h^{\mu\nu }$ in (\ref{comrel}) shows
that the constraints ${\cal H}_\mu$ close with structure functions
(depending on the field configurations) instead of structure
``constants'' (depending only on the space-time coordinates). This
fact reflects the impossibility of realizing the Lie algebra of
diffeomorphisms in phase space.

\vspace{5mm}

Acording to (\ref{delq}), in the case where the $B$ coefficients
vanish, the complete canonical generator of the gauge symmetries
in phase space is $G^H - A^{\mu_0} \phi_{\mu_0}$, with the $A$
coefficients determined by (\ref{cond-coef}). In our case this
gives (see \cite{Pons:1996pr} for a different derivation),

\begin{equation}
    G[\vec \xi] = \int d^3\!x \ \left( P_\mu \dot\xi^\mu
    + ( {\cal H}_\mu + N^\rho C^\nu_{\mu\rho} P_\nu) \xi^\mu \right)\,,
         \label{thegen}
\end{equation}

Since in all the procedure we have not undertaken any gauge
fixing, not even partially, we conclude that (\ref{thegen}) is the
general expression for the gauge generator in the entire phase
space. As it was shown in \cite{Pons:1996pr}, it describes locally
(i.e., around the identity) the diffeomorphism-induced gauge group
in phase space. In particular, its action on the configuration
variables is that of a field-dependent diffeomorphism generated by
the vector field (\ref{epsilon-xi}).

\vspace{5mm}

Finally, let us remark that the deep reason why (\ref{comrel}),
and hence the coefficients $C^\sigma_{\mu\nu}$, exhibit field
dependence is that, as we said before, the vector field $n^\mu$ in
(\ref{epsilon-xi}) is not a true vector field under
diffeomorphisms. It turns out that the second term on the r.h.s.
of (\ref{deviation}), which reflects the deviation from the vector
behavior, is directly responsible for the piece in (\ref{comrel})
that carries the $h^{\mu\nu}$ dependence, and is the ultimate
cause for the Poisson bracket algebra of the constraints ${\cal
H}_\mu$ to close with structure functions and hence to form a soft
algebra.

But, do we have still a group? Of course the gauge group, hugely
larger that the diffeomorphisms group, is always a group. The fact
that, for projectability reasons, we have chosen a basis of
generators for the gauge group in phase space that close as a soft
algebra, does not contradicts any group law for the composition of
our diffeomorphism-induced elements of the gauge group.

\section{Conclusions}

The canonical formalism for general relativity, relying on a $3+1$
decomposition, has a very specific, non-standard way of
accommodating in phase space the full diffeomorphism invariance
existing in tangent space. Being non-intrinsic, the $3+1$
decomposition is somewhat at odds with a generally covariant
formalism, and difficulties arise for this reason. The
non-projectability of some structures from tangent space to phase
space is an example of such difficulties.

Nevertheless, in the case of diffeomorphisms, a kind of compromise
is reached, and eventually we get some field-dependent
diffeomorphisms that become projectable. A basis of infinitesimal
projectable diffeomorphisms is thus obtained, and repeated
iteration -that is, exponentiation- will provide us with the
elements of the diffeomorphism-induced gauge group in phase space.

\vspace{5mm}

Let us list our main results.

\begin{enumerate}

\item

In this paper we have identified the obstruction that prevents
some infinitesimal diffeomorphisms of a generally covariant theory
like GR from being realized in phase space. A novelty of our
tratment is that this obstruction, which is an outcome of the
Noether theory of symmetries extended to gauge theories, is
identified at the level of the characterisation of the Noether
conserved conserved quantities in phase space, (\ref{cond}).

\item

We have also shown that this problem is {\sl common to the
canonical formulation of all generally covariant theories that
contain fields other than scalars}. We give a complete explanation
as to why this problem must be present. The essence of the
argument is as follows. a) A generally covariant theory containing
fields other that scalars must have first-class {\it secondary}
constraints. b) These secondary constraints appear as the Poisson
brackets of the canonical Hamiltonian with the first-class primary
constraints. c) This canonical Hamiltonian is just the Noether
conserved quantity associated with time translations -a particular
case of an arbitrary time diffeomorphism. d) Therefore the
conserved Noether quantity associated with arbitrary time
diffeomorphisms will develop $B$ terms in (\ref{cond-coef}), thus
making unavoidable the non-projectability of these time
diffeomorphisms.

\item

The adoption of field-dependent diffeomorphisms, in order to get
projectability, appears as the natural and immediate way out
within our formalism; we then recover standard formulas connecting
these diffeomorphisms with the Noether transformations realized in
phase space, which are generated -through the Poisson bracket- by
specific combinations of the constraints of the theory, all first
class. Let us mention that this method of regaining projectability
becomes more involved when other gauge fields are present. For
instance in Einstein-Yang-Mills theories, in addition to
field-dependent diffeomorphisms one must also use
\cite{Pons:1999xu} some field-dependent gauge rotations. Something
similar happens when one uses the tetrad formalism
\cite{Pons:1999ck} or the Ashtekar
\cite{ashtekar/86,ashtekar/87,ashtekar/91} complex formulation of
canonical gravity \cite{Pons:1999xt,Salisbury:1999rv}.

\item

Is is also shown that the resolution of the problem of
projectability is unavoidably linked to the fact that the
secondary constraints of GR -the so called Hamiltonian
constraints- only close under structure functions. It is a
consequence of our analysis that this is the price we must pay in
order to solve the problem of projectability of diffeomorphisms.
Going a little further we show that it is the failure of the
``vector" field orthogonal to the equal-time surfaces to behave as
a true vector under diffeomorphisms (this ``vector" is constructed
algebraically from of the components of the metric tensor) that
causes this closure under structure functions.

\item

It is worth mentioning that using the doubly contracted Bianchi
identities, (\ref{bianchi}), interpreted as the Noether
identities, (\ref{noet-id}), in the pure gravity case of GR, gives
a very efficient shortcut to obtain directly the Noether conserved
currents associated with diffeomorphism invariance.

\end{enumerate}

\vskip 6mm

{\it{\bf Acknowledgements}}

I am grateful to Jonathan Halliwell, Chris Isham and Hugh Jones
for useful discussions and comments. I am also grateful to Jos\'e
Antonio Garc{\'i}a, Xavier Gr\`acia, Donald Salisbury, and Larry
Shepley, for the many exciting interactions concerning joint work
on related subjects. This work is partially supported by MCYT FPA,
2001-3598, CIRIT, GC 2001SGR-00065, and HPRN-CT-2000-00131. I
would like to thank the Spanish ministry of education for a grant.

 \vskip 4mm




\begin{thebibliography}{99}

\bibitem{arnowitt/deser/misner/62}
R.\ Arnowitt, S.\ Deser, and C.\ W.\ Misner,
in {\it Gravitation: An Introduction to Current Research}. Edited
by L.\ Witten (John Wiley \& Sons, New York, 1962), 227--265
(1962)

\bibitem{bergmann-komar}
P. G. Bergmann and A. Komar, Int.\ J.\ Theor.\ Phys.\ {\bf 5}
(1972) 15.

\bibitem{Teitelboim:fb}
C.~Teitelboim,
{\it  In *Held, A.(Ed.): General Relativity and Gravitation,
Vol.1*, 195-225}.

\bibitem{Teitelboim:1972vw}
C.~Teitelboim, ``How Commutators Of Constraints Reflect The
Space-Time Structure,'' Annals Phys.\  {\bf 79} (1973) 542.


\bibitem{Pons:1996pr}
J.~M.~Pons, D.~C.~Salisbury and L.~C.~Shepley, ``Gauge
transformations in the Lagrangian and Hamiltonian formalisms of
generally covariant theories,'' Phys.\ Rev.\ D {\bf 55}(1997) 658
[arXiv:gr-qc/9612037].


\bibitem{Salisbury83}
D. C. Salisbury and K. Sundermeyer,
    Phys. Rev. D {\bf 27}, 740 (1983)

\bibitem{Salisbury83b}
D. C. Salisbury and K. Sundermeyer,
    Phys. Rev. D {\bf 27}, 757 (1983)

\bibitem{Pons:1999az}
J.~A.~Garcia and J.~M.~Pons, ``Rigid and gauge Noether symmetries
for constrained systems,'' Int.\ J.\ Mod.\ Phys.\ A {\bf 15}
(2000) 4681 [arXiv:hep-th/9908151].

\bibitem{dewitt} B. S. DeWitt, in ``Relativity, Groups and Topology''
Gordon and Breach, New York 1963.

\bibitem{Salisbury:1999rv}
D.~C.~Salisbury, J.~M.~Pons and L.~C.~Shepley, ``Gauge symmetries
in Ashtekar's formulation of general relativity,'' Nucl.\ Phys.\
Proc.\ Suppl.\  {\bf 88} (2000) 314 [arXiv:gr-qc/0004013].

\bibitem{Pons:1999ck}
J.~M.~Pons, D.~C.~Salisbury and L.~C.~Shepley, ``The gauge group
in the real triad formulation of general relativity,'' Gen.\ Rel.\
Grav.\  {\bf 32} (2000) 1727 [arXiv:gr-qc/9912087].

\bibitem{Pons:1999xu}
J.~M.~Pons, D.~C.~Salisbury and L.~C.~Shepley, ``Gauge
transformations in Einstein-Yang-Mills theories,'' J.\ Math.\
Phys.\  {\bf 41} (2000) 5557 [arXiv:gr-qc/9912086].

\bibitem{Pons:1999xt}
J.~M.~Pons, D.~C.~Salisbury and L.~C.~Shepley, ``Gauge group and
reality conditions in Ashtekar's complex formulation of  canonical
gravity,'' Phys.\ Rev.\ D {\bf 62} (2000) 064026
[arXiv:gr-qc/9912085].

\bibitem{Isham:1984sb}
C.~J.~Isham and K.~V.~Kuchar, ``Representations Of Space-Time
Diffeomorphisms. 1. Canonical Parametrized Field Theories,''
Annals Phys.\  {\bf 164} (1985) 288.

\bibitem{Isham:rz}
C.~J.~Isham and K.~V.~Kuchar, ``Representations Of Space-Time
Diffeomorphisms. 2. Canonical Geometrodynamics,'' Annals Phys.\
{\bf 164} (1985) 316.

\bibitem{Halliwell:1990qr}
J.~J.~Halliwell and J.~B.~Hartle,
Phys.\ Rev.\ D {\bf 43} (1991) 1170.

\bibitem{Savvidou:2001dt}
N.~Savvidou,
Class.\ Quant.\ Grav.\  {\bf 18} (2001) 3611
[arXiv:gr-qc/0104081].

\bibitem{Dirac:pj}
P.~A.~Dirac, ``Generalized Hamiltonian Dynamics,'' Can.\ J.\
Math.\  {\bf 2} (1950) 129.

\bibitem{dirac} P. A. M. Dirac,
 {\sl Can. J. Math. \bf 2} (1950) 129-148; {\it Lectures on Quantum Mechanics}
    (Yeshiva Univ.\ Press, New York, 1964)

\bibitem{Batlle:1985ss}
C.~Batlle, J.~Gomis, J.~M.~Pons and N.~Roman, ``Equivalence
Between The Lagrangian And Hamiltonian Formalism For Constrained
Systems,'' J.\ Math.\ Phys.\  {\bf 27} (1986) 2953.

\bibitem{Wald:rg}
R.~M.~Wald, ``General Relativity,'' {\it  Chicago, Usa: Univ. Pr.
( 1984) 491p}.



\bibitem{Gracia:af}
X.~Gracia and J.~M.~Pons,
J.\ Phys.\ A {\bf 25} (1992) 6357.

\bibitem{kami82} K. Kamimura, {\it Nouvo Cimento \bf B68} (1982), 33.

\bibitem{Gracia:1992pe}
X.~Gracia and J.~M.~Pons, ``Noether transformations without
conserved quantity,'' Annales Poincar\'e Phys.\ Theor.\  {\bf 61}
(1994) 315.

\bibitem{Garcia:2000yt}
J.~A.~Garcia and J.~M.~Pons, ``Lagrangian Noether symmetries as
canonical transformations,'' Int.\ J.\ Mod.\ Phys.\ A {\bf 16}
(2001) 3897 [arXiv:hep-th/0012094].

\bibitem{dompep}
J.~M.~Pons, D.~C.~Salisbury, in preparation.

\bibitem{Pons:2001vx}
J.~M.~Pons, ``Boundary conditions from boundary terms, Noether
charges and the trace  K Lagrangian in general relativity,'' Gen.\
Rel.\ Grav.\  {\bf 35} (2003) 147 [arXiv:gr-qc/0105032].

\bibitem{ashtekar/86}
A.\ Ashtekar,
    ``New variables for classical and quantum gravity,''
    Phys.\ Rev.\ Lett.\ {\bf 57}, 2244--2247 (1986)

\bibitem{ashtekar/87}
A.\ Ashtekar,
    ``New Hamiltonian formulation of general relativity,''
    Phys.\ Rev.\ {\bf D36}, 1587--1602 (1987)

\bibitem{ashtekar/91}
A.\ Ashtekar,
    {\it Lectures on Non-Perturbative Canonical Gravity}
    Notes prepared in collaboration with R.\ S.\ Tate
    (World Scientific, Singapore, 1991)
\end{thebibliography}
\end{document}